\documentclass[prb,showpacs,twocolumn,aps,a4paper]{revtex4}
\usepackage{dcolumn}
\usepackage{amsmath}
\usepackage{graphicx}
\usepackage{latexsym}
\usepackage{amsfonts}
\usepackage{amssymb}
\DeclareGraphicsExtensions{.pdf,.gif,.jpg}

\newcommand{\be}{\begin{equation}}
\newcommand{\ee}{\end{equation}}
\newcommand{\beq}{\begin{eqnarray}}
\newcommand{\eeq}{\end{eqnarray}}

\begin{document}
    
\def\bbe{\mbox{\boldmath $e$}}
\def\bbf{\mbox{\boldmath $f$}}    
\def\bg{\mbox{\boldmath $g$}}
\def\bh{\mbox{\boldmath $h$}}
\def\bj{\mbox{\boldmath $j$}}
\def\bq{\mbox{\boldmath $q$}}
\def\bp{\mbox{\boldmath $p$}}
\def\br{\mbox{\boldmath $r$}}    

\def\bone{\mbox{\boldmath $1$}}    

\def\dr{{\rm d}}

\def\tb{\bar{t}}
\def\zb{\bar{z}}

\def\tgb{\bar{\tau}}

\def\bC{\mbox{\boldmath $C$}}
\def\bG{\mbox{\boldmath $G$}}
\def\bH{\mbox{\boldmath $H$}}
\def\bK{\mbox{\boldmath $K$}}
\def\bM{\mbox{\boldmath $M$}}
\def\bN{\mbox{\boldmath $N$}}
\def\bO{\mbox{\boldmath $O$}}
\def\bQ{\mbox{\boldmath $Q$}}
\def\bR{\mbox{\boldmath $R$}}
\def\bS{\mbox{\boldmath $S$}}
\def\bT{\mbox{\boldmath $T$}}
\def\bU{\mbox{\boldmath $U$}}
\def\bV{\mbox{\boldmath $V$}}
\def\bZ{\mbox{\boldmath $Z$}}

\def\bcalS{\mbox{\boldmath $\mathcal{S}$}}
\def\bcalG{\mbox{\boldmath $\mathcal{G}$}}
\def\bcalE{\mbox{\boldmath $\mathcal{E}$}}

\def\bgG{\mbox{\boldmath $\Gamma$}}
\def\bgL{\mbox{\boldmath $\Lambda$}}
\def\bgS{\mbox{\boldmath $\Sigma$}}

\def\bgr{\mbox{\boldmath $\rho$}}

\def\a{\alpha}
\def\b{\beta}
\def\g{\gamma}
\def\G{\Gamma}
\def\d{\delta}
\def\D{\Delta}
\def\e{\epsilon}
\def\ve{\varepsilon}
\def\z{\zeta}
\def\h{\eta}
\def\th{\theta}
\def\k{\kappa}
\def\l{\lambda}
\def\L{\Lambda}
\def\m{\mu}
\def\n{\nu}
\def\x{\xi}
\def\X{\Xi}
\def\p{\pi}
\def\P{\Pi}
\def\r{\rho}
\def\s{\sigma}
\def\S{\Sigma}
\def\t{\tau}
\def\f{\phi}
\def\vf{\varphi}
\def\F{\Phi}
\def\c{\chi}
\def\w{\omega}
\def\W{\Omega}
\def\Q{\Psi}
\def\q{\psi}

\def\ua{\uparrow}
\def\da{\downarrow}
\def\de{\partial}
\def\inf{\infty}
\def\ra{\rightarrow}
\def\bra{\langle}
\def\ket{\rangle}
\def\grad{\mbox{\boldmath $\nabla$}}
\def\Tr{{\rm Tr}}
\def\Re{{\rm Re}}
\def\Im{{\rm Im}}

\title{Bound states in ab-initio approaches to quantum transport: a time-dependent 
formulation}

\author{Gianluca Stefanucci}
\email{gianluca@physik.fu-berlin.de}
\affiliation{Institut f\"ur Theoretische Physik, Freie Universit\"at Berlin, 
Arnimallee 14, D-14195 Berlin, Germany}

\date{\today}

\begin{abstract}
In this work we study the role of bound electrons in quantum transport. 
The partition-free approach by Cini is combined with time-dependent 
density functional theory (TDDFT) to calculate total currents and 
densities 
in interacting systems. We show that 
the biased electrode-device-electrode system with bound states does not evolve 
towards a steady regime. The density oscillates with 
history-dependent amplitudes and, as a consequence, the effective 
potential of TDDFT oscillates too. Such time dependence might open new conductive 
channels, an effect which is not accounted for in any steady-state approach
and might deserve further investigations.
\end{abstract}

\pacs{72.10.Bg, 73.63.-b,85.30.Mn}

\maketitle

\section{Introduction}


The progressive miniaturization of electronic devices and their future 
application in nanoscale circuitry entails the 
necessity of developing a quantum theory of transport. Such a theory
should account for the full atomistic structure of the 
contacts and the molecular device, and for the dynamical effects 
of exchange and correlation on the electron motion.

Most theoretical works on quantum transport focus on steady-state 
properties and are based on a ``contacting approach'' first 
introduced by Caroli {\em et al}.\cite{ccns.1971,cclns.1971,c.1971,ccns.1972} 
In this approach the system is separated in isolated parts in the 
remote past (left/right leads and central device) and the contacts 
between subsystems are treated as a {\em time-dependent} 
perturbation (adiabatic switching). Such a 
partition is  
crucial in the formal derivation of the Meir-Wingreen 
formula,\cite{mw.1992}
which allows for studying time-dependent phenomena and correlation effects 
in the device region but is not suitable to 
include long-range interactions and memory effects in a 
first-principle manner (see discussion below). 

The contacting approach has a severe limitation. In a typical 
experiment the whole system is contacted and in equilibrium {\em before} a  
driving external field is switched on. Thus, there is an implicit assumption 
of equivalence between  a) the initially partitioned and 
biased system once the \textit{contacts} are introduced and b) the 
whole system in equilibrium once the \textit{bias} is turned on. 
This assumption might be reasonable in the long-time limit but  
is clearly inappropriate at any finite time. As a consequence, the 
contacting approach is not suitable for describing transient regimes.
Also, the memory of initial conditions is not properly 
accounted for. In a non-interacting system
memory effects are washed out provided the local density 
of states is smooth in the device region.\cite{sa1.2004,sa2.2004,ksarg.2005} 
However, this is generally not true 
in the interacting case.\cite{sa2.2004,vsa.2006,sssbht.2006}

Recently Dhar and Sen\cite{ds.2006} have shown that
memory effects might be observed 
also in non-interacting systems with bound states.
This finding renders the contacting approach 
ambiguous since the equilibrium 
distribution of the initially isolated device is arbitrary.
It has also been shown that the one-particle 
density matrix of the unbiased system does not reduce to the standard 
equilibrium result but rather oscillates with frequencies 
$\w=\ve_{b}-\ve_{b'}$, where $\ve_{b},\ve_{b'}$ are bound-state 
energies. In order to circumvent this problem Dhar and Sen proposed 
to introduce two ``extra'' reservoirs weakly coupled to the device. 
For the unbiased system the equilibrium result is recovered in the limit 
where the ``extra'' couplings tend to zero. However, out of equilibrium 
the results depend on the limiting procedure.

In this work we use the ``partition-free approach'' by Cini\cite{c.1980} 
which is free from all above limitations. In contrast to the contacting 
approach the system is not partitioned in the remote 
past and is in equilibrium at a unique temperature and chemical 
potential (thermodynamic consistency). The initial equilibrium 
distribution of the device is unambiguous. The system is driven out of 
equilibrium by exposing the electrons to a {\em time-dependent} electric 
field. Thus, the external perturbation is a {\em local} potential and 
the partition-free approach can be combined with Time-Dependent 
Density Functional Theory\cite{rg.1984,lt.1985,vl.1998,vl.1999} (TDDFT) to calculate total 
currents and densities in interacting systems.\cite{sa2.2004,ksarg.2005} 
The use of TDDFT in quantum transport is gaining 
ground\cite{sa2.2004,ksarg.2005,ts.2001,bn.2003,bsn.2004,dvt.2004,hbf.2004,hbfts.2004,
ewk.2004,bsdv.2005,bcg.2005,qlly.2006,sssbht.2006,ly.2006,sbhdv.2007,dvda.2007} 
and several properties of the time-dependent exchange-correlation potential and 
kernel have recently been 
discussed.\cite{bmg.2005,szvdv.2005,mbg.2005,kbe.2006,bjg.2006}
 
In a previous 
work\cite{sa1.2004} we have shown how a steady current develops under 
the influence of a constant bias. For non-interacting electrons 
we also proved that the
steady current is independent of the history of the applied bias and 
agrees with the steady current calculated in the contacting 
approach. The theory has been developed assuming a smooth density of 
states in the device region. Here, we generalize the theory and include 
bound states in the description of time-dependent quantum-transport phenomena.
Our main findings are: For non-interacting electrons 
1) in the presence of bound states the total current 
and the one-particle density matrix oscillate with frequencies 
$\w=\ve_{b}-\ve_{b'}$, and the amplitude of the oscillations are 
{\em unambiguous} calculable quantities, 2) the amplitude of the 
oscillations depends on the history 
of the applied bias and on the original equilibrium, meaning that the 
long-time limit does not wash out the effects of different initial 
conditions, 3) for the unbiased system the oscillations vanish and 
all the time-dependent quantities reduce to their 
equilibrium value. Thus, in the 
partition-free approach there is no need of ``extra'' 
reservoirs to recover standard equilibrium properties. For Kohn-Sham 
electrons (TDDFT) 4) the steady-state assumption is not consistent 
with the presence of bound states, 5) it is possible to have 
self-consistent oscillatory solutions even for time-independent 
external fields, 6) density oscillations might induce oscillations in the effective 
potential of TDDFT and hence give rise to new conductive channels.
This effect can not be captured in static DFT calculations.

The plan of the paper is as follows. In Section \ref{negf} we give a 
brief introduction to the extended Keldysh formalism which 
properly accounts for the effects of initial correlations. A general 
formula for the total current is derived in Section \ref{tdi}. Such a 
formula explicitly contains the distribution function of the system 
in equilibrium and allows for working in both the contacting approach 
and the partition-free approach. In Section \ref{bsqt} 
we generalize the theory of Ref. \onlinecite{sa1.2004} to include 
bound states in quantum transport. The 
implications of our findings in a formulation based on TDDFT
are discussed in Section \ref{tddft}. In Section \ref{conc} 
we summarize the main results and draw our conclusions.

\section{The Keldysh-Green's function}
\label{negf}

\begin{figure}[htbp]
\includegraphics*[width=.48\textwidth]{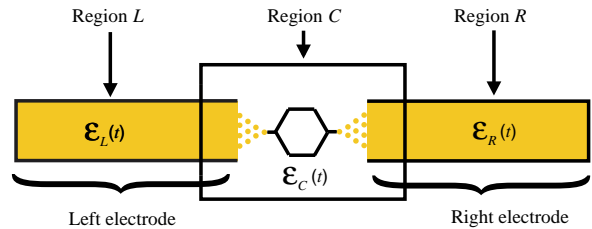}
\caption{The electrode-device-electrode system described by $\bH(t)$. 
It consists of left ($L$) and right ($R$) metallic regions coupled to a central 
scattering region $C$. The region $C$ contains the actual
molecular device and few atomic layers of the $L/R$ electrodes.}
\label{LCR}
\end{figure}

Let us consider a system of non-interacting electrons (or mean-field 
electrons or Kohn-Sham electrons) described by 
\be
\hat{H}(t)=\sum_{m,n}[\bH(t)]_{m,n}c^{\dag}_{m}c_{n},
\ee
where the matrix 
\be
\bH(t)=\left[
\begin{array}{ccc}
\bcalE_{L}(t) & \bV_{LC}(t) & 0 \\
\bV_{CL}(t) & \bcalE_{C}(t) & \bV_{CR}(t) \\
0 & \bV_{RC}(t) & \bcalE_{R}(t)
\end{array}
\right]
\label{ham}
\ee
models the electrode-device-electrode system of Fig. \ref{LCR}. (Here 
and in the following we use boldface to indicate matrices in 
one-electron labels.)

Without loss of generality we assume that the system is in equilibrium 
for negative times $t<0$. 
Letting $\m$ be the chemical potential and 
$\b$ be the inverse temperature, all equilibrium quantities
can be expressed in terms of the density matrix 
\be
\hat{\r}=e^{-\b(\hat{H}^{0}-\m\hat{N})},
\ee
with $\hat{N}=\sum_{m}c^{\dag}_{m}c_{m}$ the operator of the total number of 
particles and $\hat{H}^{0}=\hat{H}(t<0)$. The Matsubara-Green's function 
technique is a well established theory for calculating 
$\hat{\r}$-averaged quantities but is limited to equilibrium problems. 
A very powerful tool for 
dealing with non-equilibrium (as well as equilibrium) problems is 
provided by the non-equilibrium Green's function (NEGF) theory.
\cite{kb.1962,k.1965}

\begin{figure}[htbp]    
\includegraphics*[width=.45\textwidth]{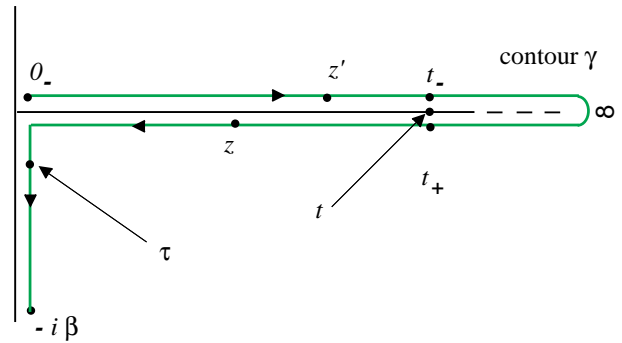}
\caption{The oriented contour $\g$ is composed by a forward and a backward branch between 
$0$ and $\inf$ and a vertical track going from $0$ to $-i\b$. 
According to the orientation the point $z$ is later than 
$z'$ and any point lying on the vertical track is later than both $z$ 
and $z'$. For any physical time $t$ we have two points $t_{\pm}$ on $\g$ at the 
same distance from the origin. In the main text we choose the Greek 
letter $\t$ for $z$ on the vertical track.}
\label{ksv}
\end{figure}
The basic quantity in NEGF is the Keldysh-Green's function
\be
[\bcalG(z;z')]_{m,n}=\frac{1}{i}
\frac{\Tr\left[{\cal T}\left\{e^{-i\int_{\g}\dr \zb \hat{H}(\zb)}
c_{m}(z)c^{\dag}_{n}(z')\right\}\right]}{   
\Tr [\hat{\r}]},
\label{kgfunc}
\ee
where $\Tr$ denotes the trace over all many-body states. In Eq. (\ref{kgfunc}) 
the integral is over the Keldysh contour $\g$ of Fig. \ref{ksv}, 
${\cal T}$ is the contour ordering operator and 
$\hat{H}(z)=\hat{H}(t)$ for $z=t_{\pm}$ while $\hat{H}(z)=\hat{H}^{0}$ 
for $z=\t$. The fermion 
operators $c_{m}(z)=c_{m}$, $c^{\dag}_{n}(z')=c^{\dag}_{n}$
do not depend on the contour variables $z$ and $z'$; the reason of the contour argument stems from 
the need of specifying their positions in the contour ordering. The 
Green's function $\bcalG$ obeys an important cyclic 
relation\cite{k.1957,ms.1959} 
\beq
\bcalG(0_{-};z')&=&-e^{\b\m}\bcalG(-i\b;z'),
\label{bc1}
\\
\label{bc2}
\bcalG(z;0_{-})&=&-e^{-\b\m}\bcalG(z;-i\b),
\eeq
which is used as boundary condition for the solution of the equation of motion
\be
\left\{i\frac{\dr}{\dr 
z}\bone-\bH(z)\right\}\bcalG(z;z')=\d(z;z')\bone,
\label{eom}
\ee
with $\bone$ the unity matrix.
The Keldysh-Green's function reduces to the Matsubara-Green's function 
for $z$ and $z'$ on the vertical track of the contour $\g$. 

In the NEGF formulation of quantum transport 
the non-local Hamiltonian 
connecting regions $L/R$ to the central region $C$ is treated 
perturbatively to all orders. This is possible by rewriting $\bH$ as 
$\bcalE+\bV$ where
\be
\bcalE=\left[
\begin{array}{ccc}
\bcalE_{L} & 0 & 0 \\
0 & \bcalE_{C} & 0 \\
0 & 0 & \bcalE_{R}
\end{array}
\right],\quad
\bV=
\left[
\begin{array}{ccc}
0 & \bV_{LC} & 0 \\
\bV_{CL} & 0 & \bV_{CR} \\
0 & \bV_{RC} & 0
\end{array}
\right],
\ee
and by introducing the diagonal Green's function $\bg$ which obeys 
the equation of motion below
\be
\left\{i\frac{\dr}{\dr z}\bone-\bcalE(z)\right\}\bg(z;z')=\d(z;z')\bone.
\label{eomd}
\ee
The diagonal $\bg$ allows us to convert Eq. (\ref{eom}) 
into a Dyson equation on $\g$ 
\be
\bcalG(z;z')=\bg(z;z')+\int_{\g}\dr\zb\; 
\bg(z;\zb)\bV(\zb)\bcalG(\zb;z'),
\label{de}
\ee
{\em provided $\bg$ obeys the same cyclic relations} 
(\ref{bc1}-\ref{bc2}) {\em as $\bcalG$}. We wish to stress 
that Eq. (\ref{de}) is an exact equation {\em only if} the contour $\g$ includes 
the forward/backward branches {\em and} the vertical 
track. As one can see by inspection, the removal of the vertical track from 
$\g$ is equivalent to set $\bV(z=\t)=\bV^{0}=0$ and hence to 
start from an initial Hamiltonian $\bH^{0}=\bcalE^{0}$, with 
$\bcalE^{0}=\bcalE(t<0)$.

To calculate the time-dependent 
total current one needs the lesser component of the Keldysh-Green's function. 
Choosing $z=t_{-}$, $z'=t'_{+}$ in Eq. (\ref{de}) and using the Langreth 
theorem\cite{l.1976,vldsavb.2005} we find
\beq
\bcalG^{<}(t;t')&=&\bg^{<}(t;t')+\int_{0}^{\inf}\dr\tb\;
\bcalG^{\rm R}(t;\tb)\bV(\tb)\bg^{<}(\tb;t')
\nonumber \\ 
&+&\int_{0}^{\inf}\dr\tb\;
\bcalG^{<}(t;\tb)\bV(\tb)\bg^{\rm A}(\tb;t')
\nonumber \\
&+&\int_{0}^{-i\b}\dr\tgb\;
\bcalG(t;\tgb)\bV(\tgb)\bg(\tgb;t'),
\label{g<1}
\eeq
where Green's functions with superscript R/A are 
retarded/advanced Green's functions. 
It is possible to show that Eq. (\ref{g<1}) is equivalent to\cite{sa1.2004}
\be
\bcalG^{<}(t;t')=\bcalG^{\rm R}(t;0)\bcalG^{<}(0;0)
\bcalG^{\rm A}(0;t').
\label{g<2}
\ee
Equation (\ref{g<2}), as opposed to Eq. (\ref{g<1}), has a simple physical 
interpretation. The lesser Green's function is completely known once we 
know how to propagate the one-electron states in time and how these 
states are populated before the system is driven out of equilibrium. The time-evolution is 
fully described by the retarded or advanced Green's functions 
$\bcalG^{\rm R,A}$, and the initial equilibrium distribution is fully 
described by 
$\bcalG^{<}(0;0)=if(\bH^{0})$, where $f(\w)=1/(e^{\b(\w-\m)}+1)$ is the Fermi 
function.

Equation (\ref{g<2}) is valid for non-interacting electrons and for mean-field 
(Hartree or Hartree-Fock) electrons; it is also valid for Kohn-Sham 
electrons since in TDDFT the electron-electron interaction is described 
by an exchange-correlation potential which is local in time.
The present work deals with non-interacting electrons (Section 
\ref{tdi} and \ref{bsqt}) and Kohn-Sham electrons (Section \ref{tddft}), 
and hence Eq. (\ref{g<2}) can always be used. Due to the non-locality in time of the 
correlation self-energy, Eq. (\ref{g<2}) is no longer 
valid beyond the Hartree-Fock approximation of many-body perturbation
theory. In this latter case the exact formula for the lesser Green's function
(which fully accounts for transient and correlation effects)  
can be found in Ref. \onlinecite{sa1.2004}.

\section{Current formula from NEGF}
\label{tdi}

The total current from region $\a=L,C,R$ can be calculated from the time 
derivative of the total number of particles in $\a$
\be
I_{\a}(t)=-e\frac{\dr}{\dr t}\Tr_{\a}\left[-i\bcalG^{<}(t;t)\right],
\ee
where $\Tr_{\a}$ denotes the trace over a complete set of one-particle 
states for region $\a$ and $e$ is the electron charge. From the 
equation of motion (\ref{eom}) the change per 
unit time of the lesser Green's function is proportional to the 
commutator $[\bcalG^{<}(t;t),\bH(t)]$. Projecting the Green's 
function onto different subregions
\be
\bcalG=\left[
\begin{array}{ccc}
\bcalG_{LL} & \bcalG_{LC} & \bcalG_{LR} \\    
\bcalG_{CL} & \bcalG_{CC} & \bcalG_{CR} \\    
\bcalG_{RL} & \bcalG_{RC} & \bcalG_{RR} \\    
\end{array}
\right],
\ee
it is straightforward to realize that
\be
I_{\a}(t)=2e\,\Re\,\Tr_{C}\left[\bcalG^{<}_{C\a}(t;t)\bV_{\a C}(t)
\right],\quad\a=L,R.
\label{current}
\ee
Equation (\ref{current}) will be our starting point for the 
calculation of the long-time limit of $I_{\a}(t)$ for different 
initial conditions and in the presence of bound states. 
Equation (\ref{current})
is valid for both non-interacting and Kohn-Sham electrons. 
In the latter case $\bcalG^{<}$ is the Kohn-Sham Green's function and 
$I_{\a}(t)$ is the total current of the Kohn-Sham 
system which, according to the Runge-Gross theorem,\cite{rg.1984} is 
the same as the total current of the real interacting system. In the 
remaining of this Section and in Section \ref{bsqt} we specialize to 
non-interacting electrons. Interacting systems will be considered in 
Section \ref{tddft}.

\subsection{The contacting approach}
\label{conta}

The contacting approach has been originally introduced by Caroli {\em et al.} 
in a series of four papers.\cite{ccns.1971,cclns.1971,c.1971,ccns.1972} 
About twenty years after it has been combined with NEGF to include the effects 
of time-dependent perturbations and short-range correlations in the scattering 
region.\cite{mw.1992,jwm.1994}
In the contacting approach regions $L$ and $R$ are
in equilibrium at the same temperature and chemical potential
and are disconnected from $C$. 
The Hamiltonian for negative times is then $\bH^{0}=\bcalE^{0}$ and the 
one-particle eigenstates of the unperturbed (and hence disconnected) 
system are strictly confined in the left region or in the right 
region or in the central scattering region. 
The equilibrium distribution of 
region $C$ can not be univocally fixed 
since $C$ is initially an isolated subsystem. This is equivalent to say 
that $\bcalE_{C}(t<0)=\bcalE_{C}^{0}$ is in principle an 
arbitrary Hamiltonian.\cite{note} We observe that this ambiguity make the transient 
current difficult to interpret.
To drive a current through the system one exposes the electrons to a 
longitudinal electric field, and at the same time switches on the contacts 
$\bV$ between $C$ and $L/R$. In a typical experimental set up regions $L$ and $R$ are bulk 
metallic electrodes, see Fig. \ref{LCR}. The dynamical formation of dipole layers at the 
$\a$-electrode interface screens the potential drop along region $\a$ and the 
total potential turns out to be uniform in the left and right bulks. 
For this reason we model the effects of a time-dependent electric 
field as a uniform shift of the energy levels, {\em i.e.}, 
$\bcalE_{\a}(t)=\bcalE^{0}_{\a}+U_{\a}(t)\bone_{\a}$, $\a=L,R$ and 
$\bone_{\a}$ the unity matrix in region $\a$.

Using Eq. (\ref{g<1}) the time-dependent current of Eq. 
(\ref{current}) can be rewritten as\cite{mw.1992}
\beq
I_{\a}(t)&=&
2e\,\Re\,\Tr_{C}\int_{0}^{\inf}\dr\tb\nonumber \\ 
&\times&\left[
\bG^{\rm R}(t;\tb)\bgS_{\a}^{<}(\tb;t)+
\bG^{<}(t;\tb)\bgS_{\a}^{\rm A}(\tb;t)
\right],
\label{ca1}
\eeq
where we have defined the so called embedding self-energy
\be
\bgS_{\a}(z;z')=\bV_{C\a}(z)\bg_{\a\a}(z;z')\bV_{\a C}(z'),
\label{se}
\ee
and introduced the short-hand notation 
\be
\bG(z;z')\equiv \bcalG_{CC}(z;z')
\ee
for the Green's function projected in region $C$. In obtaining Eq. 
(\ref{ca1}) we have taken advantage of the fact that   
$\bV(z)$ vanishes on the vertical track (isolated subsystems) and hence the 
last term in Eq. (\ref{g<1}) does not contribute.
For the same reason the embedding self-energy 
vanishes if $z$ and/or $z'$ lie on the vertical track and the 
lesser Green's function $\bG^{<}$ appearing in Eq. (\ref{ca1}) 
can then be rewritten as (see Eq. (A1) of Ref. 
\onlinecite{sa1.2004})
\beq
\bG^{<}(t;t')&=&\int_{0}^{\inf}\dr\tb\,\dr\tb'
\bG^{\rm R}(t;\tb)\bgS^{<}(\tb;\tb')\bG^{\rm A}(\tb';t')
\nonumber \\ 
&+&
\bG^{\rm R}(t;0)\bg^{<}_{CC}(0;0)\bG^{\rm A}(0;t'),
\label{g<ca}
\eeq
with $\bgS=\sum_{\a=L,R}\bgS_{\a}$ the total self-energy.
In most works on quantum transport the second term in Eq. (\ref{g<ca}) 
is neglected (see for instance Eq. (31) of 
Ref. \onlinecite{jwm.1994}). This is possible in the limit 
$t\ra\inf$ and/or $t'\ra\inf$ provided 
the local density of states (LDOS) $D_{C}$ in $C$ is a 
smooth function. However, 
$D_{C}$ is not smooth if bound states are present. 
Bound states render the 
contacting approach ambiguous because the initial equilibrium of 
region $C$ affects the behavior of the system also for $t\ra\inf$.
Moreover, as it has been recently pointed out,\cite{ds.2006} the 
time-dependent current and one-particle density matrix do not reduce 
to their equilibrium values when the perturbing electric field is set 
to zero.

In the next Section we describe the partition-free approach.
We will obtain a general expression for $I_{\a}(t)$ which  
depends explicitly on the initial equilibrium. This expression 
also allows us to switch easily between the contacting and the partition-free 
approaches.

\subsection{The partition-free approach}

The partition-free approach has been introduced by Cini\cite{c.1980} 
about a decade after the works of Caroli {\em et al.} Here the system 
is contacted and in equilibrium at a unique temperature and chemical 
potential before an external electric field is applied. The initial 
density matrix of the central region is then uniquely defined and 
transient phenomena have a direct physical interpretation.
Substituting Eq. (\ref{g<2}) into Eq. (\ref{current}) and performing 
the multiplication between Green's functions we obtain
\be
I_{\a}(t)=2e\,\Re\,\Tr_{C}\left[\bQ_{\a}(t)\right],
\quad\a=L,R
\label{current2}
\ee
with
\beq
\bQ_{\a}(t)&=&
\sum_{\g\g'=L,R}
\bcalG^{\rm R}_{C\g}(t;0)\bcalG^{<}_{\g\g'}(0;0)\bcalG^{\rm A}_{\g'\a}(0;t)
\bV_{\a C}(t)
\nonumber \\ 
&+&\sum_{\g=L,R}
\bcalG^{\rm R}_{C\g}(t;0)\bcalG^{<}_{\g C}(0;0)\bcalG^{\rm A}_{C\a}(0;t)
\bV_{\a C}(t)
\nonumber \\ 
&+&\sum_{\g'=L,R}
\bcalG^{\rm R}_{CC}(t;0)\bcalG^{<}_{C\g'}(0;0)\bcalG^{\rm A}_{\g'\a}(0;t)
\bV_{\a C}(t)
\nonumber \\ 
&+&
\bcalG^{\rm R}_{CC}(t;0)\bcalG^{<}_{CC}(0;0)\bcalG^{\rm A}_{C\a}(0;t)
\bV_{\a C}(t).
\label{qa}
\eeq
Equations (\ref{current2},\ref{qa}) are completely general. In the 
partition-free approach $\bcalG^{<}(0;0)=if(\bH^{0})=if(\bcalE^{0}+\bV^{0})$ 
while in the contacting approach 
$\bcalG^{<}(0;0)=if(\bH^{0})=if(\bcalE^{0})$ with $\bcalE^{0}_{C}$  
arbitrary. In Appendix \ref{equivalence} we prove that in the contacting approach
Eqs. (\ref{current2},\ref{qa}) are equivalent to the current formula in 
Eq. (\ref{ca1}), as it should.

In Ref. \onlinecite{sa1.2004} we have shown that by exposing the 
electrons to a constant (in time) electric field the total
current $I_{\a}(t)$ in Eqs. (\ref{current2},\ref{qa}) tends to a steady value 
$I_{\a}^{(\rm S)}$ given by (for, {\em e.g.}, $\a=L$)
\be
I_{L}^{(\rm S)}=e
\int\frac{\dr\w}{2\p}[f(\w-U_{L}^{\inf})-f(\w-U_{R}^{\inf})]
T(\w),
\label{landauer}
\ee
with $T(\w)=\Tr_{C}[
\bG^{\rm R}(\w)\bgG_{L}(\w)\bG^{\rm A}(\w)\bgG_{R}(\w)
]$, $\bgG_{\a}(\w)=-2\Im[\bgS_{\a}^{\rm R}(\w)]$, and 
$U_{\a}^{\inf}=\lim_{t\ra\inf}U_{\a}(t)$. We have also 
shown that the steady value is 
independent of the history of the applied bias (memory-loss theorem)
and of the initial equilibrium distribution of region $C$ (theorem of 
equivalence). These results were obtained for a smooth LDOS $D_{C}$. 
The presence of bound states requires a generalization of the theory. 
In the next Section we include bound states in  
time-dependent quantum transport and investigate how they 
affect the behavior of $I_{\a}(t)$ for $t\ra\inf$.

\section{Including bound states in quantum transport}
\label{bsqt}

For simplicity we consider the case of a sudden switching on of the 
bias $U_{\a}(t)=\theta(t)U_{\a}^{\inf}$ in region $\a=L,R$ (arbitrary 
time-dependent biases are considered in the next Section). 
We also specialize the discussion to a hopping Hamiltonian 
$\bV(z)=\bV^{\inf}$ that 
is constant on the forward/backward branch of the contour $\g$. However, we leave open the 
possibility of having a different constant value of $\bV(z)=\bV^{0}$ for $z$ on 
the vertical track (this allows us to deal with different approaches 
at the same time).

In quantum transport experiments the central device is typically 
connected to macroscopic metallic electrodes. 
The one-particle eigenstates $|\f_{k\a}\ket$ of 
$\bcalE_{\a}^{\inf}=\bcalE_{\a}(t\ra\inf)=
\bcalE^{0}_{\a}+U_{\a}^{\inf}\bone_{\a}$, $\a=L,R$, form a 
continuum with energy $\ve^{\inf}_{k\a}=\ve_{k\a}^{0}+U_{\a}^{\inf}
\in W_{\a}$. We define a bound state 
of the whole {\em biased} system $L+C+R$ as an eigenstate 
$|\q_{b}\ket$ of $\bH^{\inf}=\bH(t\ra\inf)=\bcalE^{\inf}+\bV^{\inf}$ with energy 
$\ve_{b}\notin W_{L}\cup W_{R}$. [For one-particle eigenstates of the 
isolated regions $\bcalE_{\a}$, $\a=L,R$, we use the Greek letter $\f$ while 
for one-particle eigenstates of the connected and biased system $\bH^{\inf}$ we use 
the Greek letter $\q$.] Below we calculate the long-time limit of 
$\bQ_{\a}$ in the presence of bound states.

\subsection{Asymptotic kernel $\bQ_{\a}$}

According to Eq. (\ref{qa}), 
the long-time limit of the kernel $\bQ_{\a}(t)$ is known once we know 
the long-time limit of $\bcalG_{CC}^{\rm R}(t;0)$, 
$\bcalG_{C\g}^{\rm R}(t;0)$, and $\bcalG_{C\a}^{\rm A}(0;t)\bV_{\a 
C}^{\inf}$, $\bcalG_{\g'\a}^{\rm A}(0;t)\bV_{\a 
C}^{\inf}$.
For any $t>0$ the retarded Green's function can be written as
\be
\bcalG^{\rm 
R}(t;0)=-i\exp\left[-i\bH^{\inf}t\right]\equiv\int\frac{\dr\w}{2\p}
\bcalG^{\rm R}(\w)e^{-i\w t},
\ee
and $\bcalG^{\rm A}(0;t)=[\bcalG^{\rm R}(t;0)]^{\dag}$. 
Exploiting the smoothness of the self-energy, 
we can use the Riemann-Lebesgue theorem to derive 
the following asymptotic behaviors
\be
\lim_{t\ra\inf}\bcalG_{CC}^{\rm R}(t;0)=-i\sum_{b}
|\q_{bC}\ket\bra\q_{bC}|e^{-i\ve_{b}t},
\label{asy1}
\ee
\beq
\lim_{t\ra\inf}\bcalG_{C\g}^{\rm R}(t;0)=
-i\sum_{b}|\q_{bC}\ket\bra\q_{bC}|\bV_{C\g}^{\inf}
\frac{e^{-i\ve_{b}t}}{\ve_{b}\bone_{\g}-\bcalE^{\inf}_{\g}}
\nonumber \\
-i\sum_{k}\bG^{\rm R}(\ve^{\inf}_{k\g})\bV^{\inf}_{C\g}
|\f_{k\g}\ket\bra\f_{k\g}|e^{-i\ve^{\inf}_{k\g}t},
\label{asy2}
\eeq
and
\be
\lim_{t\ra\inf}\bcalG_{C\a}^{\rm A}(0;t)\bV^{\inf}_{\a C}=
i\sum_{b}e^{i\ve_{b}t}|\q_{bC}\ket\bra\q_{bC}|
\bgS_{\a}^{\rm A}(\ve_{b}),
\label{asy3}
\ee
\beq
\lim_{t\ra\inf}\bcalG_{\g'\a}^{\rm A}(0;t)\bV^{\inf}_{\a C}=
i\,\d_{\g'\a}\sum_{k}e^{i\ve^{\inf}_{k\a}t}
|\f_{k\a}\ket\bra\f_{k\a}|\bV^{\inf}_{\a C}
\nonumber \\ 
+i\sum_{k}e^{i\ve^{\inf}_{k\g'}t}
|\f_{k\g'}\ket\bra\f_{k\g'}|\bV^{\inf}_{\g' C}
\bG^{\rm A}(\ve^{\inf}_{k\g'})
\bgS^{\rm A}_{\a}(\ve^{\inf}_{k\g'})
\nonumber \\
+i\sum_{b}\frac{e^{i\ve_{b}t}}{\ve_{b}\bone_{\g'}-\bcalE^{\inf}_{\g'}}
\bV^{\inf}_{\g' C}|\q_{bC}\ket\bra\q_{bC}|\bgS_{\a}^{\rm 
A}(\ve_{b}),\;\;
\label{asy4}
\eeq
where $|\q_{bC}\ket$ is the projection of eigenstate $|\q_{b}\ket$ 
onto region $C$ and $\bG^{\rm R,A}(\w)=\bcalG_{CC}^{\rm R,A}(\w)$. 

The above asymptotic results are given in terms of sum over 
bound states (discrete part) and sum over the continuum of 
states of $\bcalE_{\a}$, $\a=L,R$ (continuum part). Inserting 
them into Eq. (\ref{qa}) we obtain 
a continuum-continuum contribution, $\bQ_{\a}^{(\rm S)}$, and a 
discrete-discrete contribution, $\bQ_{\a}^{(\rm D)}$, as well as cross terms with 
discrete-continuum contributions. Taking advantage of the smoothness 
of $\bG^{\rm R,A}(\w)$ for $\w\in W_{L}\cup W_{R}$
and exploiting the Riemann-Lebesgue theorem 
it is possible to show that all cross terms vanish for $t\ra\inf$.

Let us focus on the continuum-continuum part of $\bQ_{\a}$ 
(it is straightforward to realize that only the first term in Eq. (\ref{qa}) 
can contribute to this part). 
Expanding $\bcalG^{<}(0;0)$ in Matsubara modes 
\be
\bcalG^{<}(0;0)=\frac{1}{-i\b}\sum_{n}
e^{\eta\w_{n}}\bcalG^{\rm M}(\w_{n}),
\quad\eta\ra 0
\ee
with $\bcalG^{\rm M}(\w_{n})=1/(\w_{n}\bone-\bH^{0})$ and 
$\w_{n}=\m+(2n+1)\p/(-i\b)$, we can rewrite $\bcalG^{<}_{\g\g'}(0;0)$
as
\beq
\bcalG^{<}_{\g\g'}(0;0)=if(\bcalE_{\g}^{0})\d_{\g\g'}
+\frac{1}{-i\b}\sum_{n}e^{\eta\w_{n}}\bg^{\rm M}_{\g\g}(\w_{n})
\nonumber \\ \times
\bV^{0}_{\g C}\bcalG^{\rm M}_{CC}(\w_{n})
\bV^{0}_{C\g'}\bg^{\rm M}_{\g'\g'}(\w_{n}),
\label{matzexp}
\eeq
with $\bg^{\rm M}_{\g\g}(\w_{n})=1/(\w_{n}\bone_{\g}-\bcalE^{0}_{\g})$.
We observe that in the contacting approach $\bV^{0}=0$ (regions $L,C,R$ 
isolated for negative times) and hence the second term in Eq. (\ref{matzexp}) 
vanishes. In this case, the continuum-continuum part 
is
\beq
\bQ_{\a}^{(\rm S)}=
i\int\frac{\dr\w}{2\p}f(\w-U_{\a}^{\inf})
\bG^{\rm R}(\w)\bgG_{\a}(\w)
\quad\quad\quad\quad\quad\;
\nonumber \\ 
+i\int\frac{\dr\w}{2\p}\sum_{\g}f(\w-U_{\g}^{\inf})
\bG^{\rm R}(\w)\bgG_{\g}(\w)\bG^{\rm A}(\w)\bgS_{\a}^{\rm A}(\w),
\eeq
which {\em does not depend on time}. In Appendix \ref{eqth} we prove 
that $\bQ_{\a}^{(\rm S)}$ does not depend on $\bV^{0}$ provided the hopping 
matrix elements between states in region $C$ and states 
$|\f_{k\a}\ket$, $\a=L,R$, are smooth functions of $k$. 
Thus, the memory of different initial conditions is washed out by the 
continuum of states and the final result is a steady contribution.
Below we show that this is not the case for the discrete-discrete 
part $\bQ_{\a}^{(\rm D)}$.

All four terms in Eq. (\ref{qa}) contribute to $\bQ_{\a}^{(\rm D)}$. 
After some algebra one finds the following dynamical kernel
\be
\bQ_{\a}^{(\rm D)}(t)=i\sum_{b,b'} f_{b,b'}|\q_{bC}\ket\bra\q_{b'C}|
\bgS^{\rm A}_{\a}(\ve_{b'})\,e^{-i(\ve_{b}-\ve_{b'})t},
\label{qad}
\ee
with 
\be
f_{b,b'}=\bra\q_{b}|f(\bH^{0})|\q_{b'}\ket.
\label{fbb}
\ee
The coefficient $f_{b,b'}$ is {\em the matrix element of the 
equilibrium Fermi function $f(\bH^{0})$ between bound states $b,b'$ of 
$\bH(t\ra\inf)=\bH^{\inf}$}. Equations (\ref{qad},\ref{fbb}) have been obtained 
using the bound-state contributions of the asymptotic behavior in Eqs.
(\ref{asy1}-\ref{asy4}) and the relation
\be
|\q_{b\g}\ket=\frac{1}{\ve_{b}\bone_{\g}-
\bcalE^{\inf}_{\g}}\bV_{\g C}^{\inf}|\q_{bC}\ket,
\label{bap}
\ee
with $|\q_{b\g}\ket$ the projection onto region $\g=L,R$ of the full
bound state $|\q_{b}\ket$. 

In conclusion we have
\be
\lim_{t\ra\inf}\bQ_{\a}(t)=
\bQ_{\a}^{(\rm S)}+\bQ_{\a}^{(\rm D)}(t).
\label{asyqa}
\ee
Equation (\ref{asyqa}) is completely general and is valid 
provided the Hamiltonian of region $\a=L,R$ is uniformly shifted by a 
constant potential ($\bcalE_{\a}^{\inf}=\bcalE_{\a}^{0}+
U_{\a}^{\inf}\bone_{\a}$), the contacting Hamiltonian $\bV(t)=\bV^{\inf}$ 
and the Hamiltonian for the central region $\bcalE_{C}(t)=\bcalE_{C}^{\inf}$ 
are constant in time. We observe that in deriving Eq. (\ref{asyqa}) 
no statements about $\bV^{0}$ and $\bcalE_{C}^{0}$ have been made; in 
principle, they might be different from their asymptotic values 
$\bV^{\inf}$ and $\bcalE_{C}^{\inf}$.

\subsection{Asymptotic current and one-particle density matrix}

The total current in the long-time limit is obtained by substituting 
Eq. (\ref{asyqa}) into Eq. (\ref{current2})
\be
\lim_{t\ra\inf}I_{\a}(t)=I_{\a}^{(\rm S)}+I_{\a}^{(\rm D)}(t),
\label{toti}
\ee
where $I_{\a}^{(\rm S)}$ is the steady part coming from 
$\bQ_{\a}^{(\rm S)}$ and is given in Eq. (\ref{landauer}). Taking 
into account that the imaginary part of $\bgS^{\rm A}_{\a}(\w)$ 
vanishes for $\w=\ve_{b}$, the 
dynamical part $I_{\a}^{(\rm D)}(t)$ can be written as
\be
I_{\a}^{(\rm D)}(t)=2e\sum_{b,b'}f_{b,b'}\L^{(\a)}_{b,b'}
\sin\left[(\ve_{b}-\ve_{b'})t\right],
\label{id}
\ee
with
\be
\L^{(\a)}_{b,b'}=\Tr_{C}\left[|\q_{bC}\ket\bra\q_{b'C}|
\bgS^{\rm A}_{\a}(\ve_{b'})
\right].
\ee
Proceeding along similar lines one can also prove that the 
one-particle density matrix $\bgr_{C}(t)=-i\bG^{<}(t;t)$ in region $C$ 
is the sum of steady and dynamical contributions
\be
\lim_{t\ra\inf}\bgr_{C}(t)=\bgr^{(\rm S)}_{C}+\bgr^{(\rm D)}_{C}(t),
\label{totr}
\ee
with
\be
\bgr^{(\rm S)}_{C}=\int\frac{\dr\w}{2\p}\sum_{\g}
f(\w-U_{\g}^{\inf})\bG^{\rm R}(\w)\bgG_{\g}(\w)\bG^{\rm A}(\w),
\label{rs}
\ee
and
\be
\bgr^{(\rm D)}_{C}(t)=\sum_{b,b'}f_{b,b'}|\q_{bC}\ket\bra\q_{b'C}|
\, e^{-i(\ve_{b}-\ve_{b'})t}.
\label{rd}
\ee
Equations (\ref{fbb},\ref{id},\ref{rd}) and 
the following discussion (including Section \ref{tddft}) 
are the main results of this work. 

In the contacting approach ($\bV^{0}=0$) $\bH^{0}=\bcalE^{0}$  
is a block-diagonal matrix and so is $f(\bcalE^{0})$. 
The coefficients $f_{b,b'}$ can then be rewritten as 
the sum of three terms
\be
f_{b,b'}=\bra\q_{bC}|f(\bcalE^{0}_{C})|\q_{b'C}\ket
+\sum_{\g=L,R}\bra\q_{b\g}|f(\bcalE^{0}_{\g})|\q_{b'\g}\ket.
\label{fbbca}
\ee
The first term in $f_{b,b'}$ depends on the initial equilibrium of the 
isolated central region and can not be univocally fixed (see 
discussion in Section \ref{conta}). To make contact with Ref. 
\onlinecite{ds.2006} we insert a complete set of eigenstates 
$|\f_{k\g}\ket$ for region $\g=L,R$ in the second term of 
Eq. (\ref{fbbca}). We find
\beq
\bra\q_{b\g}|f(\bcalE^{0}_{\g})|\q_{b'\g}\ket=
\sum_{k}f(\ve^{0}_{k\g})
\quad\quad\quad\quad\quad\quad\quad
\nonumber \\ 
\times
\frac{\bra\q_{bC}|\bV^{\inf}_{C\g}|\f_{k\g}\ket\bra\f_{k\g}|\bV^{\inf}_{\g C}|\q_{b'C}\ket}{
(\ve_{b}-\ve^{\inf}_{k\g})(\ve_{b'}-\ve^{\inf}_{k\g})}
\quad\quad\;
\nonumber \\ 
=
\int\frac{\dr\w}{2\p}f(\w-U^{\inf}_{\g})
\frac{\bra\q_{bC}|\bgG_{\g}(\w)|\q_{b'C}\ket}{
(\ve_{b}-\w)(\ve_{b'}-\w)},
\eeq
which agrees with the result of Dhar and Sen (see Eq. (55) of Ref. 
\onlinecite{ds.2006}). We also observe that $f_{b,b'}$ 
of Eq. (\ref{fbbca}) 
gives rise to oscillatory terms in the total current and one-particle 
density matrix which do not vanish for $U^{\inf}_{\g}=0$, $\g=L,R$. 
Thus, in the {\em unbiased} system $I_{\a}(t)$ and $\bgr_{C}(t)$ oscillate 
forever.
To solve this problem Dhar and Sen have proposed a somewhat {\em ad hoc} 
procedure which consists in introducing two extra reservoirs with 
a wide band. The bandwidth should be large enough to allow for the 
hybridization of the originally localized bound states. Eventually, 
the coupling to the extra reservoirs is removed and the standard 
equilibrium result is recovered. However, out of equilibrium this 
procedure suffers from a serious problem: The non-equilibrium 
quantities depend on how these extra couplings approach zero.

The partition-free approach is free from all the limitations described above. 
The initial equilibrium is unambiguous and  $\bcalE^{0}_{C}$
is simply given by the projection of the physical Hamiltonian onto region $C$. 
For the unbiased system the equilibrium Hamiltonian $\bH^{0}$ and 
the long-time limit Hamiltonian $\bH^{\inf}$ are the same. In this case
Eq. (\ref{fbb}) implies
\be
f_{b,b'}=\d_{b,b'}\,f(\ve_{b}),
\ee
and the dynamical contribution to the total current vanishes while 
the dynamical contribution to the one-particle density matrix 
reduces to the equilibrium contribution of bound states 
\be
\bgr^{(\rm D)}_{C}=\sum_{b}f(\ve_{b})|\q_{bC}\ket\bra\q_{bC}|,
\ee
as it should.
Out of equilibrium, both $I_{\a}(t\ra\inf)$ and $\bgr_{C}(t\ra\inf)$ oscillate 
around some steady value provided $\bH^{\inf}$ has more than one bound 
state solution. The amplitude of the oscillations are calculable 
expressions and are completely fixed by the original temperature and 
chemical potential. Non-equilibrium results are well defined.

In Appendix \ref{conservation} we also prove that Eqs. (\ref{id},\ref{rd}) 
conserve the total number of particles 
\be
\frac{\dr}{\dr t}N_{C}(t)=\frac{1}{e}\left[I_{L}(t)+I_{R}(t)\right],
\ee
where
\be
N_{C}(t)=\Tr_{C}\left[\bgr_{C}(t)\right]
\ee
is the number of particles in region $C$.

\section{Implications for TDDFT}
\label{tddft}

One of the main advantage of the partition-free approach over the 
contacting approach is that the former  
can be combined with Time-Dependent Density Functional Theory 
(TDDFT).\cite{sa1.2004,sa2.2004,ksarg.2005}
In this theory the time-dependent density of an {\em interacting} system 
can be calculated from a fictitious system of {\em non-interacting} electrons 
moving under the influence of the Kohn-Sham (KS) potential $U$.
The KS potential is the sum of the external applied potential 
$v_{\rm ext}$, the Hartree potential $v_{\rm H}$ and the exchange 
correlation potential $v_{\rm xc}$. According to the discussion of 
Section \ref{conta}, the metallic screening keeps the bulk electrodes 
in local equilibrium and we can approximate 
$U=(v_{\rm ext}+v_{\rm H}+v_{\rm xc})$ with a spatially constant 
time-dependent shift deep inside region $L/R$. The one-particle 
Hamiltonian of TDDFT is then given by Eq. (\ref{ham}) with 
$\bcalE_{\a}(t)=\bcalE_{\a}^{0}+U_{\a}(t)\bone_{\a}$, $\a= L,R$, and 
$\bV(t)=\bV^{0}$. 

Let us expose the electrons to a constant (in time) electric field 
and let us assume that the system reaches a steady state in the 
long-time limit. Then, the effective potential of the bulk electrodes
and the Hamiltonian of the central region $\bcalE_{C}$
are stationary in the distant future. We have shown that the 
steady-state assumption is consistent with 
the TDDFT equation for the total current provided the density 
of states in region $C$ is a smooth function.\cite{sa2.2004}
All history and initial-state effects are contained in $v_{\rm 
xc}({\bf r},t\ra\inf)$, meaning that two different time-dependent densities 
$n({\bf r},t)$ and $n'({\bf r},t)$ may give the same total 
current. In the Adiabatic Local 
Density Approximation (ALDA) there is no memory and the exchange-correlation potential  
depends on the instantaneous density. Hence, $v_{\rm 
xc}^{\rm ALDA}({\bf r},t\ra\inf)$ is completely known once we know 
$n({\bf r},t\ra\inf)$. The latter can be calculated 
self-consistently from Eq. (\ref{rs}). The ALDA provides a 
practical scheme to compute the current. However, such a scheme, originally 
proposed by Lang,\cite{l.1995} is obviously limited to the ALDA. Moreover, owing
to the non linearity of the problem there might be multiple 
steady-state solutions, and the absence of a minimum principle in 
out-of-equilibrium systems makes impossible to say towards which 
steady-state the electrons actually evolve.

Below we show that the 
steady-state assumption is not consistent in the presence of bound 
states. This result opens up the possibility of having self-consistent 
oscillatory solutions even for constant biases and may change 
substantially the standard 
steady-state picture. Indeed, oscillations of the effective potential in 
region $C$  give rise to new conductive channels, very much in the spirit 
of what happens in quantum pumps. 

The proof of the above statement proceeds by {\em reductio ad 
absurdum}. Let $\bH^{\inf}$ be the steady-state Hamiltonian of the 
fictitious non-interacting system. In Appendix \ref{history} we prove 
that the total current $I_{\a}(t\ra\inf)$ and one-particle density matrix
$\bgr_{C}(t\ra\inf)$ oscillate like in Eqs. (\ref{toti},\ref{totr}) 
but with new coefficients $f_{b,b'}$.
An oscillating density/current is not consistent with the 
steady-state assumption; the effective potential of 
TDDFT is a functional of the density and hence $\bH(t)$ can not be 
constant in time. We conclude that {\em a steady 
current is not compatible with the existence of bound states in the 
steady Hamiltonian} $\bH^{\inf}$.

The new coefficients 
\be
f_{b,b'}=\bra\q'_{b}|f(\bH^{0})|\q'_{b'}\ket
\label{newfbb}
\ee
depend on the history of the bias. Indeed, 
the state $|\q'_{b}\ket$ is related to the bound state 
$|\q_{b}\ket$ of $\bH^{\inf}$ by a unitary transformation 
\be
\left[\begin{array}{c}
|\q'_{bL}\ket \\
|\q'_{bC}\ket \\
|\q'_{bR}\ket
\end{array}\right]
=
\left[ \begin{array}{ccc}
e^{i\D_{L}^{\inf}}\bone_{L} & 0 & 0 \\
0 & \bM_{C} & 0 \\
0 & 0 & e^{i\D_{R}^{\inf}}\bone_{R}
\end{array}\right]
\left[\begin{array}{c}
|\q_{bL}\ket \\
|\q_{bC}\ket \\
|\q_{bR}\ket
\end{array}\right],
\label{newqb}
\ee
with 
\be
\D_{\a}^{\inf}=\lim_{t\ra\inf}
\int_{0}^{t}\dr\tb\left(U_{\a}(\tb)-U_{\a}^{\inf}\right),\quad
\a= L,R,
\ee
and $\bM_{C}$ some unitary ``memory matrix'' for region $C$. For 
constant KS potentials $\D_{\a}^{\inf}=0$ and 
$\bM_{C}=\bone_{C}$ and the coefficients $f_{b,b'}$ reduce to those 
in Eq. (\ref{fbb}).

The above results can be extended to systems of interacting electrons 
and bosons, like, e.g., phonons. We consider the system 
electrons+phonons initially in 
equilibrium and assume that the electron density is 
$v$-representable. Then, there must exist a local potential that 
reproduces the interacting electron density in a non-interacting system 
of only electrons. Such a potential defines uniquely $\bH^{0}$. Next, 
we drive the system electrons+phonons out of equilibrium by switching on a longitudinal 
electric field and we ask the question if the interacting time-dependent 
electron density can be reproduced in the non-interacting system of only 
electrons moving under the influence of a time-dependent local
potential $U$. According to the van Leeuwen theorem\cite{vl.1999} the answer is 
affirmative provided the electron-phonon interaction preserves the 
continuity equation, which is a very weak condition. Moreover, the 
local potential $U$ is unique. Again, we can conclude that if 
the Hamiltonian of the fictitious system globally converges to an asymptotic 
Hamiltonian $\bH^{\inf}$ for $t\ra\inf$, the system can not have more than one 
bound state.

We also observe that for truly non-interacting electrons the 
effective potential is equal to the external potential and does not depend 
on the density.
In this case the solution of Appendix \ref{history} is an {\em exact 
solution} and Eqs. (\ref{id},\ref{rd}), with $f_{b,b'}$ given by Eq. 
(\ref{newfbb}), can be tested against
history-dependent effects in total currents and densities. 
All history (of the applied bias) and initial-state 
dependence are contained in the coefficients $f_{b,b'}$. 

Finally, we wish to discuss a rather interesting example. 
Let us imagine to switch on and then off the bias in a system of (i) 
truly non-interacting electrons and (ii) KS electrons. 
In system (i) there is no self-consistent evolution since the
Hamiltonian is independent of the density. The asymptotic Hamiltonian 
$\bH^{\inf}$ is (trivially) equal to the initial Hamiltonian $\bH^{0}$
and hence the bound states are also eigenstates of $\bH^{0}$. Suppose that 
the bound-state energies lie in the continuum when the bias is on. 
Then, switching the 
bias off would result in a depopulation of the bound states and  
the asymptotic density matrix $\bgr_{C}(t\ra\inf)$ would, in general, differ 
from its equilibrium value. This expected result is confirmed by the 
exact solution. The system ``remembers''
that a bias has been switched on through the memory matrix $\bM_{C}$ 
and the phases $\D_{\a}^{\inf}$, $\a=L,\,R$, appearing in Eq. 
(\ref{newqb}). Equation (\ref{newqb}) defines new states $\q'_{b}\neq \q_{b}$
and according to Eq. (\ref{newfbb}) the 
coefficients $f_{b,b'}$ are no longer given by $\d_{b,b'}f(\ve_{b})$, 
as it would be in equilibrium. The situation is totally different 
in system (ii). The time evolution of interacting electrons in 
initial nonequilibrium states has been recently investigated within 
Time-Dependent Current Density Functional Theory\cite{gd.1988,v.2004}
(TDCDFT). It has been shown  
that the inclusion of memory 
effects\cite{vuc.1997} in the exchange-correlation vector potential  
leads to a dissipative KS dynamics,\cite{wu.2005,dav.2006,u.2006}
and D'Agosta and Vignale have been able 
to prove\cite{dav.2006} that the electron density evolves 
(unless forbidden by symmetry) towards the ground-state density.
Taking into account that the density of the TDDFT KS system 
is the same as the density of the TDCDFT KS system we conclude that
$\bH^{\inf}=\bH^{0}$ and, more important, that
the memory matrix $\bM_{C}={\bf 1}_{C}$ and the phases $\D_{\a=L,R}=0$ 
(mod $2\p$), independently of the history of the switching 
on/off process. Indeed, for these values of  $\bM_{C}$ and 
$\D_{\a=L,R}$ the density matrix is constant in time (and equal to the 
ground-state density matrix), and hence 
compatible with the (stationary) ground-state KS Hamiltonian.

\section{Conclusions}
\label{conc}

We have generalized the theory developed in Ref. 
\onlinecite{sa1.2004} to include bound states in quantum transport. In the 
partition-free approach the electrode-device-electrode system is 
contacted and in equilibrium at a unique temperature and chemical 
potential (thermodynamic consistency). The electrons are exposed to a 
time-dependent electric field for positive times. The external 
potential is local in space and total currents and 
density can be calculated from a fictitious system of non-interacting 
electrons, according to the Runge-Gross theorem. 

We have shown that for truly non-interacting electrons the biased 
system does not evolve toward a steady regime. The total current and density 
oscillate due to the presence of bound states. The amplitude of the 
oscillations depends on the initial temperature and chemical potential 
and on the history of the applied bias. Bound state oscillations 
might provide a probe to unveil the past of the system. 

In contrast to the contacting approach, in the partition-free 
approach
the initial equilibrium distribution of region $C$ is well defined and 
all quantities reduce to their equilibrium value by setting the 
external potential to zero. There is no need of ``extra reservoirs'' 
to equilibrate spurious bound-state oscillations.

Our findings might have interesting implications for the fictitious 
system of TDDFT. In Ref. \onlinecite{sa2.2004} we have shown that the 
time-dependent current tends to a steady value provided the  
Hamiltonian globally converges to a steady Hamiltonian and the 
density of states in region $C$ is smooth. This result 
is no longer valid in the presence of bound states. In Section \ref{tddft} 
we have shown that bound electrons in steady state regimes
lead to a contradiction: current and density would 
oscillate and, as a consequence, the effective potential of 
TDDFT would oscillate too. Steady quantities are not compatible 
with the existence of bound states. 

According to the above discussion, the effective potential of TDDFT 
might oscillate upon application of a constant bias. We expect that 
these oscillations have exponentially small amplitude deep inside the 
electrodes and are detectable only close to the molecular device.
Indeed, for truly non-interacting
electrons the amplitude of the density oscillations is proportional to the 
bound-state wavefunctions. Therefore, the KS potential is 
time-dependent in the device
region and tends exponentially to a constant (in time) deep inside the 
electrodes. In this case one can use a recently proposed
practical scheme\cite{ksarg.2005} to investigate bound-state 
dynamical effects within TDDFT. The scheme is based on the 
real-time propagation of the occupied KS orbitals and we are currently 
working at the implementation of the algorithm (which has been tested 
in one-dimensional model systems with excellent 
results\cite{ksarg.2005,vsa.2006}).
As for the case of gate voltages in electron pumping, oscillations of the effective 
potential open new conductive channels. Such an effect is completely left 
out in static DFT calculations and its possible relevance in molecular 
transport has yet to be discovered.

\begin{acknowledgments}
The author would like to acknowledge very fruitful discussions with
C. Verdozzi and also to thank S. Kurth and R. van Leeuwen for useful conversations.
This work was supported in part by the Deutsche Forschungsgemeinschaft,
DFG programme SFB658, by the EU Research and Training Network EXCITING and by 
the NANOQUANTA Network of Excellence.
\end{acknowledgments}

\appendix

\section{Equivalence between current formulas in the contacting 
approach}
\label{equivalence}

In the contacting approach regions $L,C,R$ are isolated for negative 
times and hence $\bcalG^{<}_{\g C}(0;0)=\bcalG^{<}_{C\g'}(0;0)=0$,  
$\bcalG^{<}_{\g\g'}(0;0)=\d_{\g\g'}\bg^{<}_{\g\g}(0;0)$ and
$\bcalG^{<}_{CC}(0;0)=\bg^{<}_{CC}(0;0)$. The kernel $\bQ_{\a}(t)$ 
simplifies to
\beq
\bQ_{\a}(t)&=&\sum_{\g}\bcalG^{\rm R}_{C\g}(t;0)\bg_{\g\g}^{<}(0;0)
\bcalG_{\g\a}^{\rm A}(0;t)\bV_{\a C}(t)
\nonumber \\
&+&
\bcalG^{\rm R}_{CC}(t;0)\bg^{<}_{CC}(0;0)\bcalG^{\rm A}_{C\a}(0;t)\bV_{\a C}(t).
\label{qa1}
\eeq
Extracting the retarded/advanced component of the Keldysh-Green's function from the 
Dyson equation (\ref{de}) we can express 
$\bcalG^{\rm R}_{C\g}$ in terms of $\bcalG^{\rm R}_{CC}=\bG^{\rm R}$, 
and $\bcalG_{\g\a}^{\rm A}$, $\bcalG^{\rm A}_{C\a}$ in terms of $\bcalG^{\rm 
A}_{CC}=\bG^{\rm A}$. Equation (\ref{qa1}) can then be rewritten as 
\beq
\bQ_{\a}(t)&=&\int 
\bG^{\rm R}(t;\tb)\bgS^{<}(\tb;\tb')\bG^{\rm A}(\tb';\tb'')
\bgS_{\a}^{\rm A}(\tb'';t)
\nonumber \\
&+&
\bG^{\rm R}(t;0)\bg^{<}_{CC}(0;0)\int\bG^{\rm A}(0;\tb)
\bgS_{\a}^{\rm A}(\tb;t)
\nonumber \\ 
&+&
\int \bG^{\rm R}(t;\tb)\bgS^{<}_{\a}(\tb;t),
\eeq
where the integrals (between 0 and $\inf$) are over barred time variables.
Taking into account Eq. (\ref{g<ca}) it is straightforward to realize 
that Eq. (\ref{ca1}) is actually equivalent to Eqs. 
(\ref{current2},\ref{qa}).

\section{Independence of $\bV^{0}$ in $\bQ_{\a}^{(\rm S)}$}
\label{eqth}

The continuum-continuum contribution to $\bQ_{\a}(t)$ can be written 
as $\bQ_{\a}^{(\rm S)}+\d\bQ_{\a}$. The steady value 
$\bQ_{\a}^{(\rm S)}$ originates from the first term on the right hand 
side of Eq. (\ref{matzexp}) and is independent of $\bV^{0}$. The extra 
term $\d\bQ_{\a}$ accounts for the possible effects of a 
non-vanishing coupling $\bV^{0}$ in the remote past. Here, we prove that $\d\bQ_{\a}$ vanishes in the long 
time limit. From Eq. (\ref{matzexp}) we find
\be
\d\bQ_{\a}=\frac{1}{-i\b}\sum_{n}e^{\eta\w_{n}}\d\bQ_{\a}(\w_{n}),
\ee
with
\beq
\d\bQ_{\a}(\w_{n})=
\sum_{\g}\int\frac{\dr\w}{2\p}
\frac{\bG^{\rm R}(\w)\bS_{\g}(\w)}{\w_{n}-\w+U^{\inf}_{\g}}e^{-i\w t}
\bG^{\rm M}(\w_{n})
\nonumber \\ 
\times\sum_{\g'}
\int\frac{\dr\w'}{2\p}\bS^{\dag}_{\g'}(\w')
\frac{\d_{\a\g'}+\bG^{\rm A}(\w')\bgS^{\rm A}_{\a}(\w')}
{\w_{n}-\w'+U^{\inf}_{\a}}e^{-i\w' t}.
\label{dqa}
\eeq
The matrix $\bS_{\g}(\w)$ is defined according to
\be
\bS_{\g}(\w)=\sum_{k}2\p\d(\w-\ve_{k\g}^{\inf})
\bV^{\inf}_{C\g}|\f_{k\g}\ket\bra\f_{k\g}|\bV^{0}_{\g C},
\ee
and is a smooth function of $\w$ provided the matrix elements of 
$\bV^{0}$ between states in region $C$ and states $|\f_{k\g}\ket$, 
$\g= L,R$ are smooth functions of $k$. Also, $\bS_{\g}(\w)=0$ 
for $\w\notin W_{L}\cup W_{R}$ and hence the products 
$\bS_{\g}\bG^{\rm R}$ and $\bS^{\dag}_{\g'}\bG^{\rm A}$ are smooth 
functions. At finite temperature the Matsubara frequencies have a 
finite imaginary part and the denominators in Eq. (\ref{dqa}) are well 
behaved. Exploiting the Riemann-Lebesgue theorem we conclude that both integrals 
in Eq. (\ref{dqa}) vanish for $t\ra\inf$, meaning that $\d\bQ_{\a}=0$.

\section{Conservation of the particle number}
\label{conservation}

The time derivative of the number of particles in region $C$ can be 
easily calculated from Eq. (\ref{rd})
\be
\frac{\dr}{\dr t}N_{C}(t)=-\sum_{b,b'}
(\ve_{b}-\ve_{b'})f_{b,b'}S_{b,b'}
\sin\left[(\ve_{b}-\ve_{b'})t\right],
\label{nc}
\ee
with $S_{b,b'}$ the overlap between bound states $b,b'$ in $C$
\be
S_{b,b'}=\Tr_{C}\left[\,|\q_{bC}\ket\bra\q_{b'C}|\,\right].
\ee
Let us now consider the expression for the total current. The sum 
$I_{L}^{(\rm S)}+I_{R}^{(\rm S)}$ of steady contributions vanishes. 
Taking into account that $f_{b,b'}$ is symmetric under the exchange 
of $b$ and $b'$, and that the real part of the advanced self-energy is a 
Hermitian matrix we can safely replace $\L_{b,b'}^{(\a)}$ with
\be
\L_{b,b'}^{(\a)}=\frac{1}{2}\Tr_{C}\left[|\q_{bC}\ket\bra\q_{b'C}|
\left(\bgS^{\rm A}_{\a}(\ve_{b'})-\bgS^{\rm A}_{\a}(\ve_{b})\right)
\right]
\label{nlam}
\ee
in the expression for the dynamical contribution to $I_{\a}(t)$. The 
difference between self-energies at different bound-state energies is
\be
\bgS^{\rm A}_{\a}(\ve_{b'})-\bgS^{\rm A}_{\a}(\ve_{b})=
\bV^{\inf}_{C\a}\frac{(\ve_{b}-\ve_{b'})}{[\ve_{b'}\bone_{\a}-\bcalE^{\inf}_{\a}]
[\ve_{b}\bone_{\a}-\bcalE^{\inf}_{\a}]}\bV_{\a C}^{\inf},
\ee
and hence Eq. (\ref{nlam}) becomes
\be
\L_{b,b'}^{(\a)}=\frac{(\ve_{b}-\ve_{b'})}{2}
\Tr_{\a}\left[\,|\q_{b\a}\ket\bra\q_{b'\a}|\,\right],
\ee
where we have used Eq. (\ref{bap}). Exploiting the orthonormality 
relation 
\be
\sum_{\a=L,C,R}\Tr_{\a}\left[\,|\q_{b\a}\ket\bra\q_{b'\a}|\,\right]=
\d_{b,b'},
\ee
we find
\be
\sum_{\a=L,R}\L_{b,b'}^{(\a)}=-\frac{(\ve_{b}-\ve_{b'})}{2}S_{b,b'}.
\ee
It is now straightforward to realize that the sum $I_{L}^{(\rm 
D)}+I_{R}^{(\rm D)}$ of dynamical contributions is equal to the 
change per unit time of the number of particle in region $C$ [which 
is given in Eq. (\ref{nc})].

\section{Long-time limit of TDDFT}
\label{history}

We have already shown in Section \ref{bsqt} that for a sudden 
switching on,  
$\bH(t>0)=\bH^{\inf}$, $\forall t>0$, 
the total current and one-particle density matrix oscillate 
in the long-time limit. Let us denote
with a bar ($\bar{\bg}$, $\bar{\bcalG}$, and $\bar{\bgS}$) 
Green's functions and self-energies corresponding to this case. 
The asymptotic expressions of 
$\bar{\bcalG}_{CC}^{\rm R}(t;0)$, $\bar{\bcalG}_{C\g}^{\rm R}(t;0)$, and 
$\bar{\bcalG}_{C\a}^{\rm A}(0;t)\bV_{\a C}^{0}$, 
$\bar{\bcalG}_{\g'\a}^{\rm A}(0;t)\bV_{\a C}^{0}$ are given in
Eqs. (\ref{asy1}-\ref{asy4}). Below, we will find a relation between 
the asymptotic behavior of $\bcalG$ and $\bar{\bcalG}$, with $\bcalG$ 
the Green's function of the TDDFT Hamiltonian. 

For arbitrary time-dependent potentials 
in region $\a=L,R$, the retarded self-energy can be expressed in 
terms of $\bar{\bgS}^{\rm R}$ as
\be
\bgS^{\rm R}_{\a}(t;t')=e^{-i\D_{\a}(t)}\bar{\bgS}^{\rm R}_{\a}(t;t')
e^{i\D_{\a}(t')},
\ee
with
\be
\D_{\a}(t)=\int_{0}^{t}\dr\tb\left(U_{\a}(t)-U_{\a}^{\inf}\right).
\ee
The quantity $\D_{\a}(t)$ depends on the history of the applied bias. 
We assume $U_{\a}(t)$ to approach $U_{\a}^{\inf}$ rapidly enough that 
$\D_{\a}(t)$ converges to some value $\D_{\a}^{\inf}$ for $t\ra\inf$.
Then,
\be
\lim_{t,t'\ra\inf}\bgS^{\rm R}_{\a}(t;t')=\bar{\bgS}^{\rm R}_{\a}(t;t').
\label{asyself}
\ee

The above identity allows us to fix the asymptotic behavior of the 
Green's function in region $C$. From the equation of motion we have
\beq
\left\{i\frac{\dr}{\dr t}\bone_{C}-\bcalE_{C}(t)\right\}
\bcalG_{CC}^{\rm R}(t;0)\quad\quad\quad\,
\nonumber \\ 
-\int\dr\tb \,\bgS^{\rm R}_{\a}(t;\tb)
\bcalG_{CC}^{\rm R}(\tb;0)
=\d(t).
\label{gcc}
\eeq
For large $t$ we can replace $\bcalE_{C}(t)$ 
with its asymptotic value $\bcalE_{C}^{\inf}$. Moreover, taking into 
account that the self-energy vanishes when the separation between its 
time arguments goes to infinity we can replace $\bgS^{\rm 
R}_{\a}(t;\tb)$ with $\bar{\bgS}^{\rm R}_{\a}(t;\tb)$, in accordance 
with Eq. (\ref{asyself}). Then, Eq. (\ref{gcc}) reduces to the 
equation obeyed by $\bar{\bcalG}_{CC}^{\rm R}(t\ra\inf;0)$, meaning that
\be
\lim_{t\ra\inf}\bcalG_{CC}^{\rm R}(t;0)=
\lim_{t\ra\inf}\bar{\bcalG}_{CC}^{\rm R}(t;0)\bM_{C}^{\dag},
\label{gccasy}
\ee
where $\bM_{C}^{\dag}$ is some unitary matrix that accounts for the history of 
the applied bias (memory effects). From Eq. (\ref{gccasy}) and the equation of motion 
for $\bcalG^{\rm A}_{C\a}(0;t)$ one can also prove that
\be
\lim_{t\ra\inf}\bcalG^{\rm A}_{C\a}(0;t)\bV^{0}_{\a C}=
\lim_{t\ra\inf}\bM_{C} \bar{\bcalG}^{\rm A}_{C\a}(0;t)\bV^{0}_{\a C}.
\label{gcaasy}
\ee

Also the asymptotic behavior of $\bcalG^{\rm R}_{C\g}$ can be 
calculated from the equation of motion. We have
\beq
\left\{i\frac{\dr}{\dr t}\bone_{C}-\bcalE_{C}(t)\right\}
\bcalG_{C\g}^{\rm R}(t;0)-\bV^{0}_{C\g}\bg^{\rm R}_{\g\g}(t;0)
\nonumber \\
-\int\dr\tb \,\bgS^{\rm R}_{\a}(t;\tb)
\bcalG_{C\g}^{\rm R}(\tb;0)=0.
\eeq
Taking the limit $t\ra\inf$ and exploiting the relation
$\bg_{\g\g}^{\rm R}(t\ra\inf;0)=e^{-i\D_{\g}^{\inf}}\bar{\bg}^{\rm 
R}_{\g\g}(t\ra\inf;0)$ we obtain
\be
\lim_{t\ra\inf}\bcalG^{\rm R}_{C\g}(t;0)=e^{-i\D_{\g}^{\inf}}
\lim_{t\ra\inf}\bar{\bcalG}^{\rm R}_{C\g}(t;0).
\label{gcgasy}
\ee
In a similar way one can prove that
\be
\lim_{t\ra\inf}\bcalG^{\rm A}_{\g'\a}(0;t)\bV_{\a C}^{0}
=e^{i\D_{\g'}^{\inf}}
\lim_{t\ra\inf}\bar{\bcalG}^{\rm A}_{\g'\a}(0;t)\bV_{\a C}^{0}.
\label{ggaasy}
\ee

Substituting these results [Eqs. (\ref{gccasy}-\ref{gcaasy}) and 
Eqs. (\ref{gcgasy}-\ref{ggaasy})] in Eqs. (\ref{current2}-\ref{qa}) one 
can calculate the asymptotic behavior of the time-dependent 
total current. Proceeding along the same line which leads to 
Eq. (\ref{toti}) one can show that $I_{\a}(t)$ is again given by the 
sum of the steady-state value $I_{\a}^{({\rm S})}$ of Eq. (\ref{landauer}) 
and the dynamical contribution $I^{({\rm D})}_{\a}(t)$ of Eq. (\ref{id}). 
However, the coefficients $f_{b,b'}$ are in general different from those 
in Eq. (\ref{fbb}). After some algebra one readily finds that the new 
coefficients $f_{b,b'}$ can be expressed as in Eqs. (\ref{newfbb}-\ref{newqb}).

The time-dependent one-particle density matrix also has the same
analytic form of Eqs. (\ref{totr}-\ref{rd}) but with the same new 
coefficients $f_{b,b'}$ of Eqs. (\ref{newfbb}-\ref{newqb}).


\begin{thebibliography}{10} 
    
\bibitem{ccns.1971}
C. Caroli, R. Combescot, P. Noz\`ieres, and D. Saint-James, 
J. Phys. C {\bf 4},  916  (1971).

\bibitem{cclns.1971}
C. Caroli, R. Combescot, D. Lederer, P. Noz\`ieres, and D. Saint-James, 
J. Phys. C {\bf 4},  2598  (1971).

\bibitem{c.1971}
R. Combescot, 
J. Phys. C 4, 2611 (1971).

\bibitem{ccns.1972}
C. Caroli, R. Combescot, P. Nozieres, and D. Saint-James, 
J. Phys. C. {\bf 5}, 21 (1972).    

\bibitem{mw.1992}
Y. Meir and Ned S. Wingreen,
Phys. Rev. Lett. {\bf 68}, 2512 (1992).

\bibitem{sa1.2004}
G. Stefanucci and C.-O. Almbladh, 
Phys. Rev. B {\bf 69}, 195318 (2004).

\bibitem{sa2.2004}
G. Stefanucci and C.-O. Almbladh, 
Europhys. Lett. {\bf 67},  14  (2004).

\bibitem{ksarg.2005}
S. Kurth, G. Stefanucci, C.-O. Almbladh, A. Rubio and E. K. U. Gross, 
Phys. Rev. B {\bf 72}, 035308 (2005).

\bibitem{vsa.2006}
C. Verdozzi, G. Stefanucci and C.-O. Almbladh,
Phys. Rev. Lett. {\bf 97}, 046603 (2006).

\bibitem{sssbht.2006}
C. G. Sanchez, M. Stamenova, S. Sanvito, D. R. Bowler, A. P. Horsfield, T. N. Todorov,
J. Chem. Phys. {\bf 124}, 214708 (2006). 

\bibitem{ds.2006}
A. Dhar and D. Sen,
Phys. Rev. B {\bf 73}, 085119 (2006).

\bibitem{c.1980}
M. Cini, 
Phys. Rev. B {\bf 22},  5887  (1980).

\bibitem{rg.1984}
E. Runge and E.~K.~U. Gross, 
Phys. Rev. Lett. {\bf 52},  997  (1984).

\bibitem{lt.1985}
T. Li and P. Tong, 
Phys. Rev. A {\bf 31}, 1950 (1985).

\bibitem{vl.1998}
R. van Leeuwen, 
Phys. Rev. Lett. {\bf 80}, 1280 (1998).

\bibitem{vl.1999}
R. van Leeuwen, 
Phys. Rev. Lett. {\bf 82}, 3863 (1999).

\bibitem{ts.2001}
J. K. Tomfohr, and O. F. Sankey,
Phys. Stat. Sol. B {\bf 226}, 115 (2001).

\bibitem{bn.2003}
R. Baer, and D. Neuhauser,
Int. J. Quantum Chem. {\bf 91}, 524 (2003).

\bibitem{bsn.2004}
R. Baer, T. Seideman, and D. Neuhauser,
J. Chem. Phys. {\bf 120}, 3387 (2004).

\bibitem{dvt.2004}
M. Di Ventra, and T. N. Todorov,
J. Phys.: Condens. Matter {\bf 16}, 8025 (2004).

\bibitem{hbf.2004}
A. P. Horsfield, D. R. Bowler, and A. J. Fisher,
J. Phys.: Condens. Matter {\bf 16}, L65 (2004).

\bibitem{hbfts.2004}
A. P. Horsfield, D. R. Bowler, A. J. Fisher, T. N. Todorov, C. G. Sanchez,
J. Phys.: Condens. Matter {\bf 16}, 8251 (2004).

\bibitem{ewk.2004}
F. Evers, F. Weigend, and M. Koentopp,
Phys. Rev. B {\bf 69}, 235411 (2004).

\bibitem{bsdv.2005}
N. Bushong, N. Sai, and M. Di Ventra,
NanoLett. {\bf 5}, 2569 (2005).

\bibitem{bcg.2005}
K. Burke, R. Car, and R. Gebauer,
Phys. Rev. Lett. {\bf 94}, 146803 (2005)

\bibitem{qlly.2006}
X. Qian, J. Li, X. Lin, and S. Yip,
Phys. Rev. B {\bf 73}, 035408 (2006).

\bibitem{ly.2006}
X.-Q. Li and Y. Yan,
Phys Rev B {\bf 75}, 075114 (2007).

\bibitem{sbhdv.2007}
N. Sai, N. Bushong, R. Hatcher, and M. Di Ventra,
cond-mat/0701634.

\bibitem{dvda.2007}
M. Di Ventra, and R. D'Agosta,
cond-mat/0702272.

\bibitem{bmg.2005}
P. Bokes, H. Mera, and R. W. Godby,
Phys. Rev. B {\bf 72}, 165425 (2005). 

\bibitem{szvdv.2005}
N. Sai, M. Zwolak, G. Vignale, and M. Di Ventra,
Phys. Rev. Lett. {\bf 94}, 186810 (2005).

\bibitem{mbg.2005}
H. Mera, P. Bokes, and R. W. Godby,
Phys. Rev. B {\bf 72}, 085311 (2005). 

\bibitem{kbe.2006}
M. Koentopp, K. Burke, and F. Evers,
Phys. Rev. B {\bf 73}, 121403(R) (2006).

\bibitem{bjg.2006}
P. Bokes, J. Jung, and R. W. Godby,
cond-mat/0604317.

\bibitem{kb.1962}
L.~P. Kadanoff and G. Baym, 
{\em Quantum Statistical Mechanics} 
(W. A. Benjamin, Inc. New York, 1962).

\bibitem{k.1965}
L. V. Keldysh, 
Zh. Eskp. Teor. Phys. {\bf 47}, 1515 (1964);
Sov. Phys. JETP {\bf 20} 1018 (1965).

\bibitem{k.1957}
R. Kubo, 
J. Phys. Soc. Jpn. {\bf 12}, 570 (1957).

\bibitem{ms.1959}
P.~C. Martin and J. Schwinger, 
Phys. Rev. {\bf 115}, 1342 (1959).

\bibitem{l.1976}
D.~C. Langreth,  in {\em Linear and Nonlinear Electron Transport in Solids},
edited by J.~T. Devreese and E. van Doren (Plenum, New York, 1976), pp.\
3--32.

\bibitem{vldsavb.2005}
R. van Leeuwen, N. E. Dahlen, G. Stefanucci, C.-O. Almbladh, and U. von Barth,
Lect. Notes Phys. {\bf 706}, 33 (2006).

\bibitem{jwm.1994}
A.-P. Jauho, N.~S. Wingreen, and Y. Meir, 
Phys. Rev. B {\bf 50},  5528 (1994).

\bibitem{note}
Any non-interacting density matrix in region $C$ can be written as 
$\hat{\r}_{C}=\exp[-\hat{K}_{C}]$, with 
$\hat{K}_{C}=\sum_{i,j}[\bK_{C}]_{i,j}c^{\dag}_{i}c_{j}$ 
some hermitian operator. This is equivalent to choose $\bcalE_{C}^{0}$
according to $\b(\bcalE_{C}^{0}+\m\bone_{C})=\bK_{C}$.    

\bibitem{l.1995}
N.~D. Lang, Phys. Rev. B {\bf 52},  5335  (1995).

\bibitem{gd.1988}
S. K. Ghosh, and A. K. Dhara,
Phys. Rev. A {\bf 38}, 1149 (1988).

\bibitem{v.2004}
G. Vignale,
Phys. Rev. B {\bf 70}, 201102(R) (2004).

\bibitem{vuc.1997}
G. Vignale, C. A. Ullrich, and S. Conti,
Phys. Rev. Lett. {\bf 79}, 4878 (1997).

\bibitem{wu.2005}
H. O. Wijewardane, and C. A. Ullrich,
Phys. Rev. Lett. {\bf 95}, 086401 (2005).

\bibitem{dav.2006}
R. D'Agosta and G. Vignale,
Phys. Rev. Lett. 96, 016405 (2006).

\bibitem{u.2006}
C. A. Ullrich, 
J. Chem. Phys. {\bf 125}, 234108 (2006).


\end{thebibliography}
\end{document}